\documentclass[runningheads]{llncs}
\usepackage[T1]{fontenc}
\usepackage{graphicx}
\usepackage{algorithm}
\usepackage{amsmath}
\usepackage{algpseudocode}
\usepackage{placeins}
\usepackage{xcolor}
\usepackage{hyperref}
\usepackage{subcaption}
\usepackage{amssymb}
\usepackage{esvect}

\begin{document}
\title{Proportional Sensitivity in Generative Adversarial Network (GAN)-Augmented Brain Tumor Classification Using Convolutional Neural Network}
\author{
Mahin Montasir Afif\inst{1}\orcidID{0009-0003-8499-2624} 
\and Abdullah Al Noman\inst{1}\orcidID{0009-0005-9843-7247} 
\and K. M. Tahsin Kabir\inst{2}\orcidID{0009-0009-5060-5060}
\and Md. Mortuza Ahmmed\inst{1}\orcidID{0000-0002-4735-571X}
\and Md. Mostafizur Rahman\inst{1}\orcidID{0000-0002-3788-1118}
\and Mufti Mahmud\inst{3,4,5}\orcidID{0000-0002-2037-8348}
\and Md. Ashraful Babu\inst{6}\orcidID{0000-0003-2144-6610}
}
\authorrunning{M. M. Afif et al.}
%
\institute{
American International University-Bangladesh, Dhaka-1229, Bangladesh
\email{22-46573-1@student.aiub.edu, 22-46609-1@student.aiub.edu, mortuza@aiub.edu, mostafiz.math@aiub.edu}\\
\and
Asian University of Bangladesh, Dhaka, Bangladesh\\
\email{tahsin@aub.ac.bd}\\
\and
Information and Computer Science Department, King Fahd University of Petroleum and Minerals (KFUPM), Saudi Arabia\\
\email{mufti.mahmud@kfupm.edu.sa}\\
\and
SDAIA-KFUPM Joint Research Center for AI, King Fahd University of Petroleum and Minerals, Dhahran, 31261, Saudi Arabia\\
\and
Interdisciplinary Research Center for Bio Systems and Machines, King Fahd University of Petroleum and Minerals, Dhahran, 31261, Saudi Arabia\\
\and
Independent University, Bangladesh, Dhaka - 1229, Bangladesh
\email{ashraful388@iub.edu.bd}
}
\maketitle              
\begin{abstract}
Generative Adversarial Networks (GAN) have shown potential in expanding limited medical imaging datasets. This study explores how different ratios of GAN-generated and real brain tumor MRI images impact the performance of a CNN in classifying healthy vs. tumorous scans. A DCGAN was used to create synthetic images which were mixed with real ones at various ratios to train a custom CNN. The CNN was then evaluated on a separate real-world test set. Our results indicate that the model maintains high sensitivity and precision in tumor classification, even when trained predominantly on synthetic data. When only a small portion of GAN data was added, such as 900 real images and 100 GAN images, the model achieved excellent performance, with test accuracy reaching 95.2\%, and precision, recall, and F1-score all exceeding 95\%. However, as the proportion of GAN images increased further, performance gradually declined. This study suggests that while GANs are useful for augmenting limited datasets especially when real data is scarce, too much synthetic data can introduce artifacts that affect the model's ability to generalize to real world cases.

\end{abstract}

\section{Introduction}
\begin{figure}
    \centering
    \includegraphics[width=0.8\linewidth]{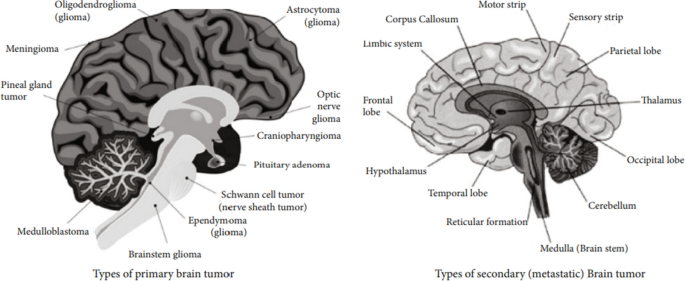}
    \caption{Types of tumor in human brain \cite{mijwil2025smart}}
    \label{fig:types}
\end{figure}

Brain tumors are among the most severe and life-threatening neurological conditions, with both benign and malignant types posing significant health risks. As shown in Figure \ref{fig:types} (gliomas, meningiomas, pituitary tumors, and metastatic cancers), these neoplasms vary considerably in their morphology and clinical impact. According to the IARC GLOBOCAN 2020 database, brain and central nervous system (CNS) tumors resulted in over 207,000 deaths globally in 2020 \cite{iarc_globocan_2020}, with glioblastomas (Grade IV gliomas) being particularly aggressive and difficult to treat \cite{who_cancer_2023}. This mortality burden underscores the urgent need for automated, accurate diagnostic systems. Deep learning approaches combining real and synthetic medical imaging data are emerging as critical tools for early tumor detection and precise classification across these diverse tumor types.
\label{method}
\FloatBarrier

\subsection{Convolutional Neural Networks for Brain Tumor Classification}
Convolutional Neural Networks (CNNs) \cite{lecun1998gradient} are a type of deep learning model designed to process image data by automatically extracting relevant visual features from basic shapes and edges to complex structures like tumor patterns. A study by \cite{DEEPAK2019103345} achieved 94.58\% accuracy using transfer learning with AlexNet/VGG-16 on T1-weighted MRI images, demonstrating its effectiveness even with limited data. \cite{saeedi2023mri} Developed a 2D CNN that reached 96.47\% accuracy and near-perfect AUC-ROC (0.99–1), outperforming traditional methods while maintaining clinical practicality.
\cite{nazir2024customized} Developed a customized CNN achieving 98.67\% validation accuracy on BR35H dataset (3,060 MRIs), integrated with SHAP, LIME, and Grad-CAM for interpretability. The model demonstrated exceptional performance (F1-score: 98.5\%), setting a new benchmark for reliable AI-assisted tumor detection. Another recent work by \cite{aamir2024optimized} presents a hyperparameter-optimized CNN for brain tumor detection, achieving 97\% accuracy across multiple MRI datasets. A paper by \cite{eity2025dggxnethybriddeeplearning} proposed a hybrid model which achieved 91.33\% test accuracy for multiclass brain disease classification including brain tumor. These studies collectively demonstrate CNNs' growing reliability and clinical utility for brain tumor detection across diverse datasets and architectures.

\subsection{Generative Adversarial Networks in Brain Tumor Classification}
Generative Adversarial Networks (GANs) \cite{goodfellow2014generative} have emerged as a transformative approach for addressing data scarcity in medical imaging through their unique generator-discriminator framework. These networks synthesize highly realistic tumor representations, enabling enhanced training of diagnostic models. Pioneering work by \cite{mukherkjee2022brain} demonstrated this potential through their AGGrGAN architecture, which combined DCGAN and WGAN variants with style transfer techniques to achieve exceptional structural similarity (SSIM=0.83) on BraTS 2020 datasets, while preserving critical tumor features for downstream analysis. Recent advancements have significantly expanded GAN applications in neuroimaging. The DIR-GAN framework \cite{karpakam2025enhanced} represents a major innovation, integrating attention mechanisms with adaptive noise filtering (AMBF) and chaotic optimization to achieve unprecedented 98.9\% classification accuracy. This hybrid approach demonstrates superior multi-modal performance across MRI, X-ray, and FigShare datasets through its novel multi-scale feature extraction and residual connections. For clinical deployment, \cite{neelima2022optimal} developed an optimized pipeline combining DeepMRSeg segmentation with GAN classification, utilizing a bio-inspired SPO algorithm to achieve 90\% segmentation accuracy and 91.7\% diagnostic precision. Their work established practical benchmarks for automated tumor diagnosis in real-world settings. Comparative studies by \cite{akbar2024brain} provide critical insights into synthetic data utility, demonstrating that GAN-generated images can support 80-90\% of real data performance in segmentation tasks. Their comprehensive evaluation of Progressive GAN and StyleGAN variants, alongside diffusion models, highlights both the potential and current limitations of synthetic data approaches, particularly regarding dataset size requirements.

In this study, we present a comprehensive framework that combines a DCGAN \cite{radford2015unsupervised}  architecture with a purpose-built CNN classifier. Our approach offers a systematic evaluation of model performance across progressively blended datasets, ranging from fully real to increasingly synthetic brain MRI samples providing one of the first in-depth analyses of the impact of GAN-generated data on tumor detection accuracy. The proposed hybrid system introduces three key innovations:
\begin{itemize}
    \item A DCGAN generator capable of synthesizing high-quality tumor features, enabling the creation of realistic MRI samples that mimic genuine pathological patterns.
    
    \item A CNN classifier optimized to learn from both real and synthetic images, ensuring reliable performance across hybrid training data.

    \item A rigorous evaluation protocol to assess the effect of synthetic data inclusion on classification performance, highlighting the trade-offs and benefits at varying real-to-GAN ratios.
\end{itemize}

\section{Methodology}

\label{method}
\begin{figure}
    \centering
    \includegraphics[width=1\linewidth]{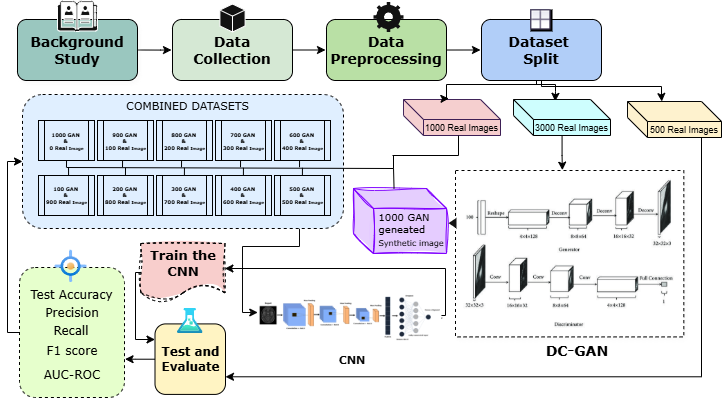}
    \caption{Methodology Overview: A comprehensive outline of the systematic procedures and techniques utilized throughout the study to ensure accurate and reliable outcomes}
    \label{fig:method}
\end{figure}
\FloatBarrier
This section outlines the end-to-end methodology, from data preparation to model evaluation, based on the workflow shown in Figure~\ref{fig:method}. The process includes synthetic image generation using DCGAN, constructing datasets with different real/synthetic ratios, training a CNN, and evaluating its performance.

\subsection{Data Collection and Preprocessing}

\subsubsection{Dataset Description:}
This study utilized the BR35H dataset \cite{br35h-::-brain-tumor-detection-2020_dataset}, which contains 3,060 brain MRI images divided into three folders: \texttt{yes} containing 1500 tumorous data, \texttt{no} containing 1500 non-tumorous data and \texttt{pred} containing 60 images. For our task, we utilized two folders \texttt{yes} and \texttt{no} containing a total of 3000 images of MRI scans.

\subsubsection{Preprocessing and Splitting:}
The dataset was partitioned into 1,000 images for further training, 3,000 for GAN training, and 500 reserved as an independent test set. All images were resized to $64 \times 64$ pixels and center-cropped to focus on the brain region, reducing background noise and artifacts. Images were normalized to $[-1,1]$ by applying:

\begin{equation}
x_{\text{norm}} = \frac{x - 0.5}{0.5}
\label{eq:normalization}
\end{equation}

where $x$ is the pixel intensity, scaled between 0 and 1. Although the images are grayscale, they were converted to RGB format to match the input size required by typical CNNs.

The preprocessing pipeline implemented \textbf{transforms.Compose} from PyTorch, including resizing, cropping, tensor conversion, and normalization. A custom dataset class applied these transforms dynamically, and data was loaded with a batch size of 64 using a DataLoader with shuffling enabled. This consistent preprocessing was applied to both real and GAN-generated images to ensure uniform input distribution for CNN training.

\subsection{DCGAN}
To augment the dataset, a Deep Convolutional Generative Adversarial Network (DC-GAN) was employed. It synthesized 1000 artificial brain MRI images. These generated samples were evenly split with 500 depicting tumor-bearing brains and 500 representing non-tumorous cases.

The DC-GAN framework is composed of two neural networks: a generator $G_\theta$ which learns to produce realistic images from noise vectors, and a discriminator $D_\phi$ which attempts to distinguish between real MRI images and those generated by the generator. The training follows a minimax optimization paradigm:

\begin{equation}
\min_{\theta} \max_{\phi} \; \mathcal{V}(D_\phi, G_\theta) = \mathbb{E}_{x \sim P_{\text{real}}}[\log D_\phi(x)] + \mathbb{E}_{\mathbf{z} \sim P_{\mathbf{z}}}[\log(1 - D_\phi(G_\theta(\mathbf{z})))]
\label{eq:dcgan_objective}
\end{equation}

Here, $P_{\text{real}}$ denotes the distribution of real MRI scans, and $P_{\mathbf{z}}$ is the prior over the input noise vector $\mathbf{z}$, typically sampled from a standard normal distribution.

The discriminator’s objective is to maximize the probability of correctly identifying real versus generated data, leading to the following loss function during training:

\begin{equation}
\mathcal{L}_{\text{disc}} = -\left( \mathbb{E}_{x \sim P_{\text{real}}}[\log D_\phi(x)] + \mathbb{E}_{\mathbf{z} \sim P_{\mathbf{z}}}[\log(1 - D_\phi(G_\theta(\mathbf{z})))] \right)
\label{eq:disc_loss_alt}
\end{equation}

Conversely, the generator is optimized to fool the discriminator into classifying generated images as real, with its loss defined as:

\begin{equation}
\mathcal{L}_{\text{gen}} = -\mathbb{E}_{\mathbf{z} \sim P_{\mathbf{z}}}[\log D_\phi(G_\theta(\mathbf{z}))]
\label{eq:gen_loss_alt}
\end{equation}

This adversarial training process enables the generator to learn the underlying distribution of the MRI dataset, allowing it to produce high-fidelity synthetic brain scans that contribute to model robustness and generalizability.

\subsection{Dataset Construction}
A total of 11 types of datasets were constructed to evaluate the performance of the CNN across various ratios, which is a key aspect of this investigation. Each dataset combination consisted of exactly 1000 images to ensure a fair and unbiased evaluation.

\begin{table}[htbp]
\centering
\caption{Dataset Combinations of Real and GAN-Generated MRI Images for training the CNN}
\label{tab:real_gan_combinations}
\begin{tabular}{|c|c|c|c|}
\hline
\textbf{Real : GAN Ratio} & \textbf{Real Images} & \textbf{GAN Images} & \textbf{Total Images} \\
\hline
100:0 & 1000 & 0    & 1000 \\
90:10 & 900  & 100  & 1000 \\
80:20 & 800  & 200  & 1000 \\
70:30 & 700  & 300  & 1000 \\
60:40 & 600  & 400  & 1000 \\
50:50 & 500  & 500  & 1000 \\
40:60 & 400  & 600  & 1000 \\
30:70 & 300  & 700  & 1000 \\
20:80 & 200  & 800  & 1000 \\
10:90 & 100  & 900  & 1000 \\
0:100 & 0    & 1000 & 1000 \\
\hline
\end{tabular}
\end{table}
\FloatBarrier

\subsection{CNN Architecture and Training Details} 

\label{cnn}
\begin{figure}
    \centering
    \includegraphics[width=1\linewidth]{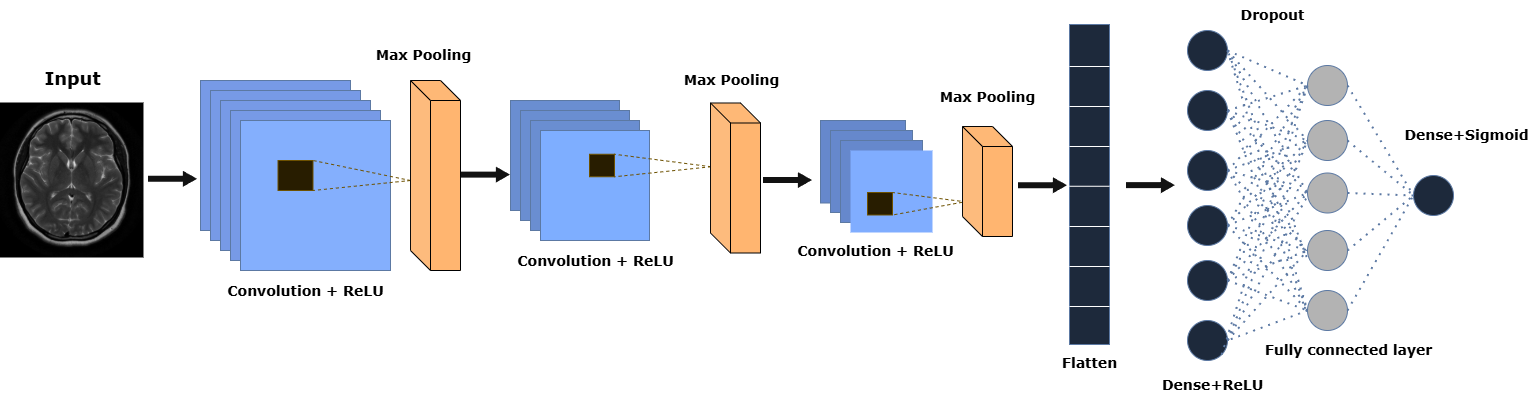}
    \caption{A tailored convolutional neural network designed and optimized specifically for the task}
    \label{fig:cnn}
\end{figure}
\FloatBarrier

A custom Convolutional Neural Network (CNN) shown in figure \ref{fig:cnn} was developed for binary classification of brain MRI scans. The model includes three convolutional layers with filter sizes of $3 \times 3$, using 32, 64, and 128 filters respectively, each followed by ReLU activation and max-pooling. This is followed by a flatten layer, a dense layer with 128 units and dropout for regularization, and a final output layer with a single neuron and sigmoid activation for binary output. The total number of trainable parameters is approximately 3.29 million.

The model was trained using binary cross-entropy loss, defined for a mini-batch of $M$ samples as:

\begin{equation}
\mathcal{L}_{\text{bin}} = -\frac{1}{M} \sum_{j=1}^{M} \left[ t_j \log(p_j) + (1 - t_j)\log(1 - p_j) \right]
\label{eq:binary_loss}
\end{equation}

where $t_j$ is the true label and $p_j$ is the predicted probability. Training used the Adam optimizer with a learning rate of $1 \times 10^{-4}$. During training, the data ratio was consistently maintained at 800:200. For testing, the same independent dataset was used throughout, consisting of 1,000 real images equally distributed across two classes: healthy and tumor.
Here is the algorithm of our custom CNN.
\begin{algorithm}[H]
\caption{CNN Training Procedure for Binary Brain MRI Classification}
\label{alg:cnn_training}
\begin{algorithmic}[1]
\Require Dataset $D = \{(x_i, y_i)\}_{i=1}^N$, learning rate $\eta$, batch size $B$, epochs $E$
\Ensure Trained CNN model $f_\theta$
\State Initialize CNN model parameters $\theta$ randomly
\State Set optimizer as Adam with learning rate $\eta$
\For{epoch $= 1$ to $E$}
    \State Shuffle dataset $D$
    \For{each mini-batch $\{(x_b, y_b)\}_{b=1}^B$ in $D$}
        \State Forward pass: compute predictions $\hat{y}_b = f_\theta(x_b)$
        \State Compute binary cross-entropy loss:
        \[
        \mathcal{L} = -\frac{1}{B} \sum_{b=1}^{B} \left[ y_b \log(\hat{y}_b) + (1 - y_b) \log(1 - \hat{y}_b) \right]
        \]
        \State Backward pass: compute gradients $\nabla_\theta \mathcal{L}$
        \State Update parameters: $\theta \leftarrow \theta - \eta \cdot \nabla_\theta \mathcal{L}$
    \EndFor
    \State Optionally evaluate on validation set and store metrics
\EndFor
\State \Return Trained model $f_\theta$
\end{algorithmic}
\end{algorithm}

\subsection{Evaluation Metrics}

All models were evaluated on a separate real-world test set using the following metrics \cite{bishop2006pattern} \cite{ahmmed2025modelmediatedstackedensembleapproach}:

\begin{equation}
\text{Accuracy} = \frac{C_1 + C_2}{C_1 + C_2 + I_1 + I_2}
\label{eq:custom_accuracy}
\end{equation}

\begin{equation}
\text{Precision} = \frac{C_1}{C_1 + I_1}
\label{eq:custom_precision}
\end{equation}

\begin{equation}
\text{Recall} = \frac{C_1}{C_1 + I_2}
\label{eq:custom_recall}
\end{equation}

\begin{equation}
\text{F1-Score} = 2 \times \frac{\text{Precision} \cdot \text{Recall}}{\text{Precision} + \text{Recall}}
\label{eq:custom_f1}
\end{equation}

where:
\begin{itemize}
    \item $C_1$ = correctly classified positive cases (true positives),
    \item $C_2$ = correctly classified negative cases (true negatives),
    \item $I_1$ = incorrectly classified negative cases as positive (false positives),
    \item $I_2$ = incorrectly classified positive cases as negative (false negatives).
\end{itemize}

\vspace{0.3em}
Additionally, the Area Under the Receiver Operating Characteristic Curve (AUC-ROC) was used to measure the model's ability to distinguish between classes.
Formally, the AUC can be expressed as:

\begin{equation}
\text{AUC} = \frac{1}{M_+ \cdot M_-} \sum_{i=1}^{M_+} \sum_{j=1}^{M_-} \mathbb{I}(s_i > s_j)
\label{eq:auc_mannwhitney}
\end{equation}

where:
\begin{itemize}
    \item $M_+$ and $M_-$ are the number of positive and negative samples, respectively,
    \item $s_i$ and $s_j$ are the predicted scores for a positive and a negative sample,
    \item $\mathbb{I}(s_i > s_j)$ is the indicator function which returns 1 if $s_i > s_j$, else 0.
\end{itemize}

\section{Result and Discussion}
\label{results}
This section presents a comprehensive analysis of the CNN model’s performance when trained using varying proportions of GAN-generated and real-world brain tumor MRI images. The goal was to assess the impact of synthetic image augmentation on classification accuracy, sensitivity, and robustness in detecting brain tumors from real test data.

We conducted a series of experiments where the model was trained using combinations of real and GAN-generated images in different ratios, ranging from purely real data (0\% GAN) to fully synthetic data (100\% GAN). A total of 11 different training configurations were tested, each maintaining a consistent split of 800 images for training and 200 for validation with testing always performed on the same real-world dataset.

\subsection{Synthetic Data Generation using DCGAN}
\label{sec:results}

The results of training the Deep Convolutional Generative Adversarial Network (DCGAN) over 1000 epochs are presented in this section. The network was trained on real-world brain MRI scans to generate a synthetic dataset containing 500 tumor and 500 non-tumor brain images. Throughout the training process, generator and discriminator losses were closely monitored to assess convergence behavior and model stability.

\subsubsection{GAN Training Performance:}
\begin{figure}[htbp]
    \centering
    \begin{subfigure}[b]{0.49\textwidth}
        \centering
        \includegraphics[width=\linewidth]{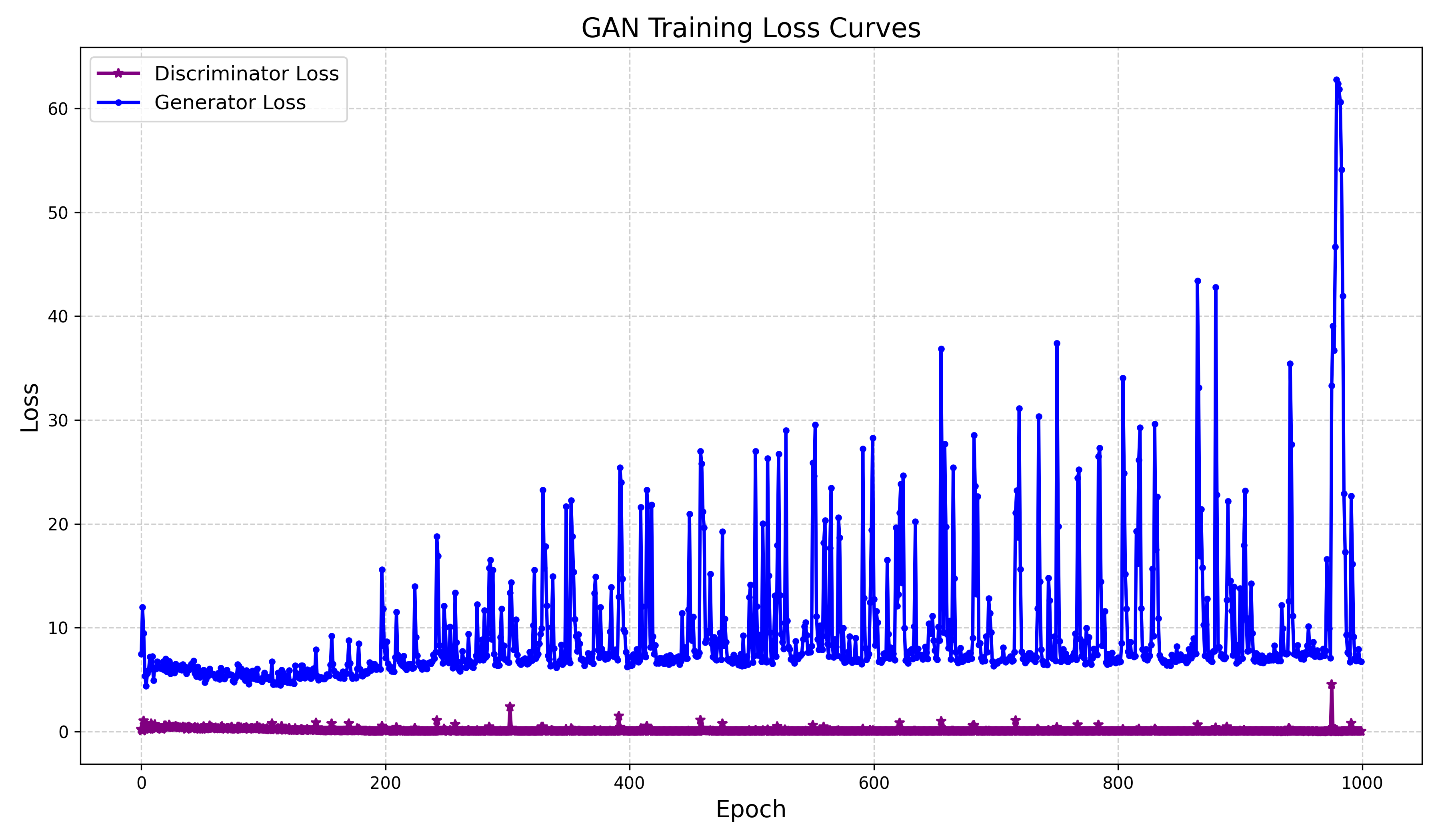}
        \caption{Generator and Discriminator Losses for Tumor Image Generation}
        \label{fig:tumor_gan_loss_curve}
    \end{subfigure}
    \hfill
    \begin{subfigure}[b]{0.49\textwidth}
        \centering
        \includegraphics[width=\linewidth]{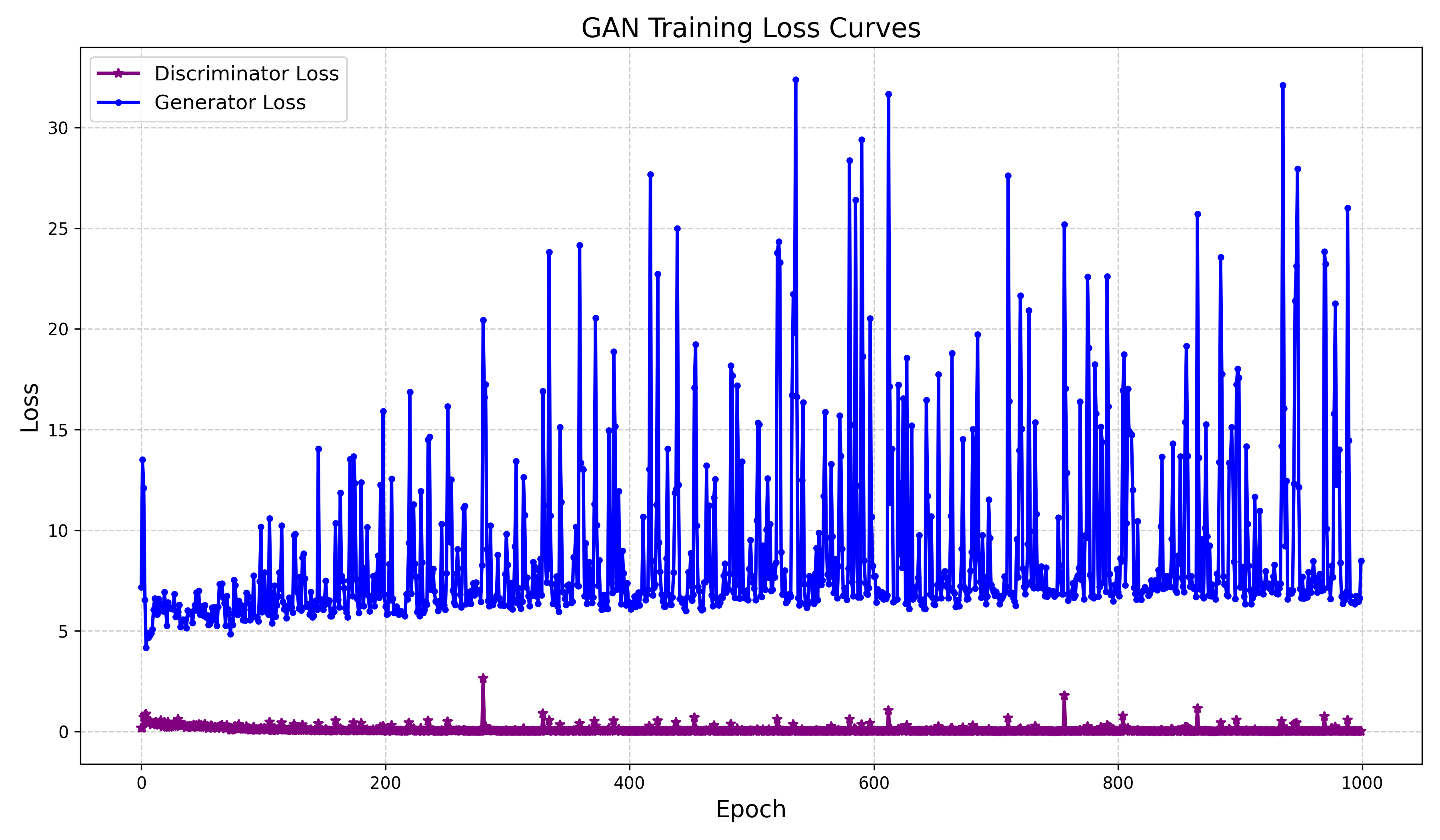}
        \caption{Generator and Discriminator Losses for Non-Tumor Image Generation}
        \label{fig:non_tumor_gan_loss_curve}
    \end{subfigure}
    
    \caption{Tumor and Non-Tumor GAN Training Curves Across 1000 Epochs}
    \label{fig:gan_loss_curves}
\end{figure}

The generator and discriminator were trained using adversarial loss optimization with both networks utilizing a learning rate of $2 \times 10^{-4}$. The training process was analyzed independently for the tumor and non-tumor image generation tasks to better understand convergence behavior across different classes. Figure~\ref{fig:tumor_gan_loss_curve} shows the generator and discriminator losses during tumor image synthesis, while Figure~\ref{fig:non_tumor_gan_loss_curve} illustrates the same for non-tumor image generation.

\FloatBarrier
\subsubsection{Final Synthetic Dataset:}

\begin{figure}[htbp]
    \centering
    \includegraphics[width=0.8\linewidth]{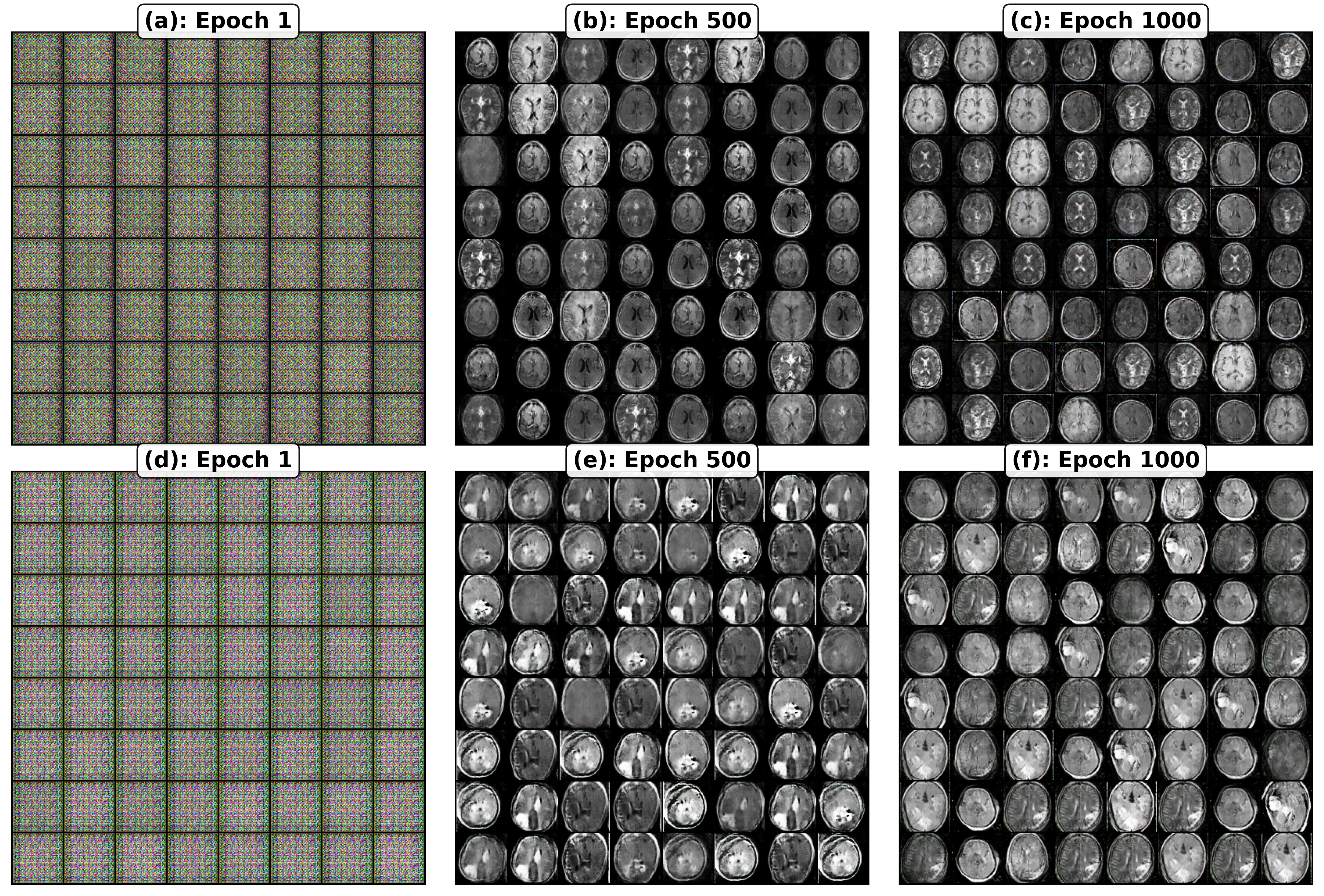}
    \caption{Progressive visual results produced by the GAN at various training stages}
    \label{fig:epoch}
\end{figure}

After 1000 epochs, the generator was used to produce a total of 1000 synthetic brain MRI images: 500 labeled as tumor and 500 as non-tumor. Visual inspection of generated images at different stages of training revealed progressive improvement in image quality and realism. Figure~\ref{fig:epoch} displays representative samples from the GAN output:
Subfigures \textbf{(a)}, \textbf{(b)}, and \textbf{(c)} show non-tumor (healthy brain) images generated at Epoch 1, Epoch 500, and Epoch 1000 respectively. Subfigures \textbf{(d)}, \textbf{(e)}, and \textbf{(f)} illustrate the progression of tumor brain image generation across the same epochs. As shown in figure~\ref{fig:epoch}, the synthetic images closely resemble real MRIs by the final training epochs.
\FloatBarrier

\subsection{Training and Validation Performance Across Various GAN-Real Data Ratios using Custom CNN}

The performance of the custom CNN model across different real-to-GAN training ratios is summarized in Table~\ref{tab:data_distribution_comparison}. As the proportion of GAN-generated images increased, a gradual decline in accuracy, precision, recall, and F1-score was observed which indicates the model's sensitivity to domain shift. This trend highlights that while GANs can support training when used sparingly, over-reliance on synthetic data leads to diminished generalization on real-world brain tumor detection tasks.

\begin{table}[htbp]
\caption{Performance of CNN across varying real and GAN data proportions}
\label{tab:data_distribution_comparison}
\centering
\scriptsize
\begin{tabular}{|l|c|c|c|c|c|}
\hline
\textbf{Data Distribution} & \textbf{Accuracy (\%)} & \textbf{Precision (\%)} & \textbf{Recall (\%)} & \textbf{F1-Score (\%)} & \textbf{AUC} \\
\hline
0\% GAN, 100\% Real & 94.70 & 93.40 & 96.20 & 94.78 & 0.98 \\
10\% GAN, 90\% Real & \textbf{95.20} & \textbf{94.31} & \textbf{96.20} & \textbf{95.25} & \textbf{0.98} \\
20\% GAN, 80\% Real & 94.10 & 96.62 & 91.40 & 93.94 & 0.98 \\
30\% GAN, 70\% Real & 92.90 & 92.81 & 93.00 & 92.91 & 0.98 \\
40\% GAN, 60\% Real & 89.80 & 89.33 & 90.40 & 89.86 & 0.95 \\
50\% GAN, 50\% Real & 88.40 & 86.78 & 90.60 & 88.65 & 0.95 \\
60\% GAN, 40\% Real & 86.70 & 91.24 & 81.20 & 85.93 & 0.94 \\
70\% GAN, 30\% Real & 85.70 & 82.51 & 90.60 & 86.37 & 0.93 \\
80\% GAN, 20\% Real & 79.70 & 85.96 & 71.00 & 77.77 & 0.88 \\
90\% GAN, 10\% Real & 76.60 & 73.84 & 82.40 & 77.88 & 0.85 \\
100\% GAN, 0\% Real & 65.40 & 73.05 & 48.80 & 58.51 & 0.71 \\
\hline
\end{tabular}
\end{table}

\begin{figure}[htbp]
\centering
\begin{minipage}{0.49\linewidth}
    \centering
    \includegraphics[width=\linewidth]{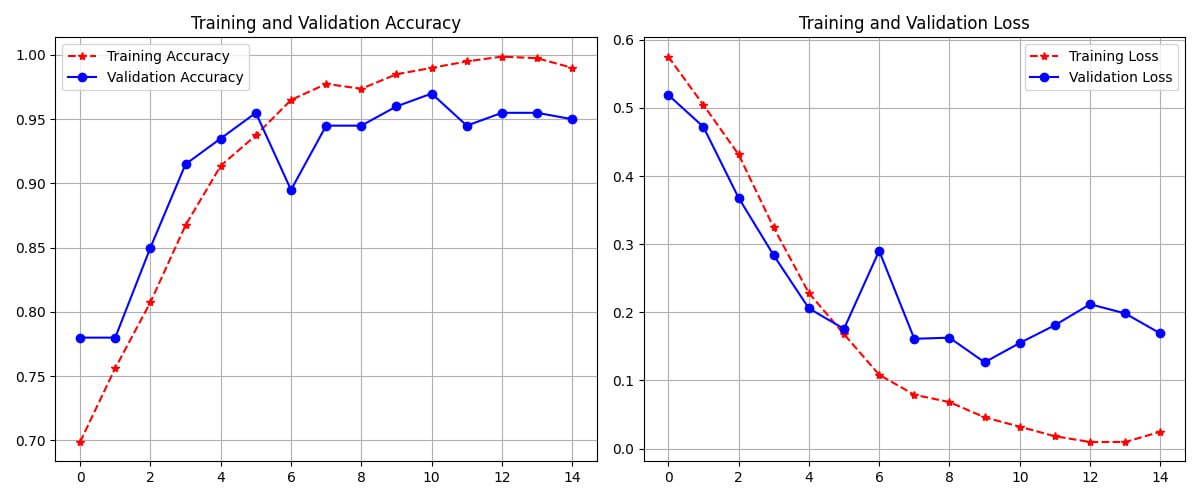}
    \caption{Training and validation accuracy and loss curves on the best-performing dataset proportion.}
    \label{fig:trainloss}
\end{minipage}
\hspace{0.02\linewidth} 
\begin{minipage}{0.4\linewidth}
    \centering
    \includegraphics[width=\linewidth]{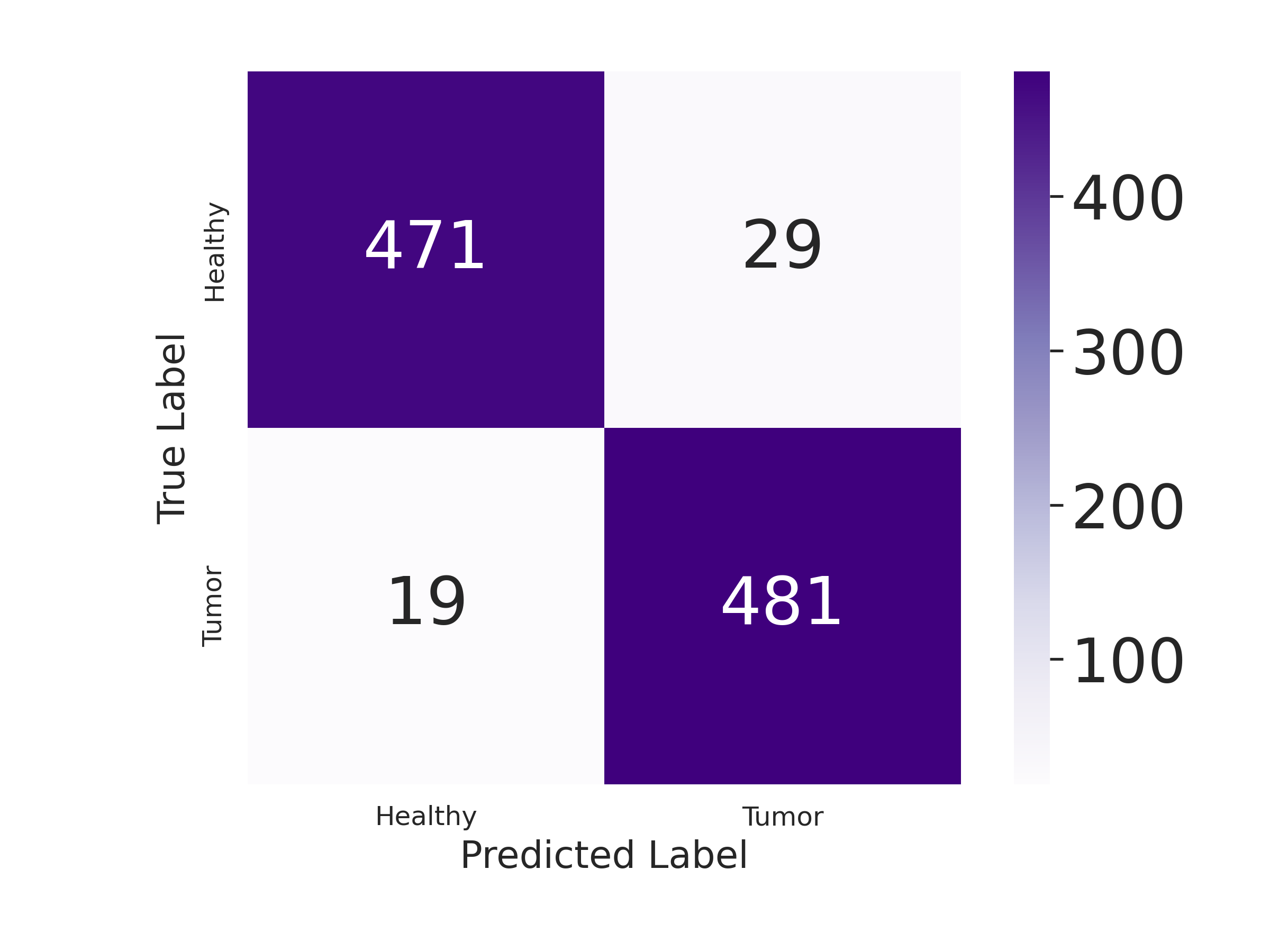}
    \caption{Confusion matrix on the best-performing dataset proportion.}
    \label{fig:confusion}
\end{minipage}
\end{figure}

\subsection{Discussion }
Overall, through this study we observe a consistent trend: as the proportion of real data decreases and synthetic (GAN) data increases, the classification performance gradually declines across all metrics, Accuracy, Precision, Recall, F1-Score, and AUC. The best performance was achieved using a combination of 10\% GAN and 90\% real data, with an accuracy of 95.20\%, F1-Score of 95.25\%, and AUC of 0.98. Figures~\ref{fig:trainloss} and~\ref{fig:confusion} illustrate the training and validation curves, and the confusion matrix, respectively.
 This suggests that a small inclusion of GAN-generated data may enhance generalization by introducing slight variability, acting as a regularizer. However, as the synthetic ratio increases further (especially beyond 50\%), the model's performance starts to deteriorate significantly. The lowest results were observed in the 100\% GAN case, with accuracy dropping to 65.40\% and AUC to 0.71 which indicates that the GAN-generated data alone may lack sufficient diversity or realism to fully replace real images. These results support the hypothesis that while GAN-generated images can supplement real data effectively, they cannot fully substitute real data in sensitive medical imaging tasks. A balanced ratio with real data as the majority is essential for optimal classification performance.
\section{Conclusion}
\label{conclusion}

This innovative study demonstrates the critical role of data augmentation using GANs in advancing automated medical image classification systems. In the domain of medical imaging where labeled datasets are often limited due to privacy, cost, or rarity synthetic data generation offers a promising solution for improving model training and generalization. This experiment provides a comprehensive overview of how varying proportions of real and GAN-generated brain MRI images affect CNN performance, revealing that while small amounts of GAN data can enhance learning without compromising accuracy, over-reliance on synthetic samples leads to degradation in precision and reliability.

Our findings underscore the sensitivity of CNN-based classifiers to domain shifts between synthetic and real-world data, highlighting the importance of careful integration of GANs in medical AI development. This work not only validates the usefulness of GAN-augmented training but also sets a foundation for future research into more realistic and controllable synthetic image generation techniques such as diffusion models and domain-adapted GANs.
By systematically analyzing how GAN-generated images impact CNN training, this study contributes valuable insight into the practical application of synthetic data in healthcare AI pipelines, paving the way for safer, more scalable deep learning solutions in medical diagnostics.



%

\bibliographystyle{splncs04}
\bibliography{ref}

\end{document}